# Cluster Identification and Characterization of Physical Fields

Zhang Guang-Cai[+], Xu Ai-Guo, Lu Guo, Mo Ze-Yao

National Key Laboratory of Computational Physics, Institute of Applied Physics and Computational Mathematics, Beijing 100088

**The description of complex configuration is a difficult issue. We present a powerful technique for cluster identification and characterization. The scheme is designed to treat with and analyze the experimental and/or simulation data from various methods. Main steps are as follows. We first divide the space using face or volume elements from discrete points. Then, combine the elements with the same and/or similar properties to construct clusters with special physical characterizations. In the algorithm, we adopt administrative structure of hierarchy-tree for spatial bodies such as points, lines, faces, blocks, and clusters. Two fast search algorithms with the complexity $\ln N$ are realized. The establishing of the hierarchy-tree and the fast searching of spatial bodies are general, which are independent of spatial dimensions. Therefore, it is easy to extend the skill to other fields. As a verification and validation, we treated with and analyzed some two-dimensional and three-dimensional random data.**

Spatial hierarchical tree; fast search; complex configuration; dynamic physical fields; cluster identification

Complex configuration and dynamic physical fields are ubiquitous in weapon-physics, astrophysics, plasma-physics, and material-physics. Those structures and their evolutions are characterizing properties of the corresponding physical systems. For example, the interface instability makes significant constrains on the design of inertial confinement fusion(ICF) device[1], shock wave and jet-flow in high energy physics are common phenomena[2], distributions of clouds and nebulae are very concerned issues of astrophysics[3,4,5], clusters and filaments occur in the interaction of high-power laser and plasma[6], structures of dislocation band determine the material softening in plastic deformation of metals[7]. These structures are also keys to understand the multi-scale physical processes. Laws in small-scale determines the growth, the change and the interactions of stable structures in larger-scale. Description of evolution of the stable structures provides constitutive relation for larger-scale modeling. Since lacking periodicity, symmetry, spatial uniformity or pronounced correlation, the identification and characterization of these structures have been challenging for years.

*Supported by National Natural Science Foundation of China (Grant Nos. 10102010 and 10775018), Science Foundations of Laboratory of Computational Physics and China Academy of Engineering Physics (Grant Nos. 2009A0102005 and 2009B0101012).
[+]Corresponding author.E-mail：zhang_guangcai@iapcm.ac.cn

Existing methods for analyzing complex configurations and dynamic fields include the linear analysis of small perturbation of background uniform field, characteristic analysis of simple spatial distribution of physical fields, etc. These methods are lack of quantitative description of characteristics of the physical domain. For example, the size, the shape, the topology, the circulation and the integral of physical quantity. Therefore, it is difficult to trace the evolution of the characteristic region or the background. For example, the laws of growth and decline, or the exchange between them.

The difficulties in characteristic analysis are twofold. The first is how to define the characteristic region. The second is how to describe it. The former involves the control equations of the physical system. The latter is related to recovering the geometric structure from discrete points. In recent years, cluster analysis techniques [8, 9] in data mining have found extensive applications in identification and law-exploration of targets. They mainly concern the schemes for data classification. Physists concern more the nature underlying these structures. Recovering characteristics domains can be attributed to the construction of spatial geometry. The key point is how to connect the related discrete points. The Delaunay grid[10, 11] has an excellent spatial neighbor relationship. In this work we use the Delaunay triangle or tetrahedron as the foundamental geometrical element.

The designs of data structure and algorithm for fast searching are core issues in the field of computer software engineering [12, 13]. Using tree structure to manage spatial discrete data has obvious advantages in memory usage and fast searching. In the fields of celestial evolution and galaxy formation, space hierarchy tree (SHT) is widely used to manage distribution of space particles [14, 15] so that the forces on particles and mass distribution of galaxies can be fastly calculated.

In this paper we use the spatial hierarchy tree to manage objects in n-dimensional space. Two general adaptive fast searching algorithms are presented. As applications, Delaunay division and cluster structure construction in two- and three-dimensional spaces are performed. Following parts of the paper are organized as below. Section 1 describes the SHT managing structure and its construction. Section 2 outlines two fast-searching methods based on the SHT. Section 3 describes the fast methods to construct Delaunay tetrahedron or triangle based on the two fast-searchers. Cluster construction algorithm and its analysis are presented in section 4. Finally, in section 5 we make a short summary and briefly explore potential applications of the SHT.

**1 SHT management structure**

SHT has been successfully applied to the management and indexing of spatial data points [16], but has not yet applied to more complex spatial objects, such as lines, surfaces, bodys, clusters, etc. In this paper we use the SHT to manage objects with spatial location, shape and size. The basic idea is as follows. For a system in n-dimensional space, we design a n-dimensional cube to contain the system; and then divide this cube in each dimension into two parts to form $2^n$ sub-cubes; only retain the cubes with objects inside; continue to decompose each cube until the required resolution is reached; put the objects (points, lines, surfaces, bodies) into appropriate cube according to their locations and sizes; existing cubes are connected together, according to their belonging relationships, to form a 'spatial hierarchy tree'. Each cube is named a 'branch'. Its child-cubes are named 'sub-branches' and its parent-cube is called the 'trunk'. The largest cube is named the 'root'. Figure 1 (Figure 2) is schematic for the SHT management structure of two-dimensional (three-dimensional) discrete points. Due to the uncertainty of number of 'sub-branches' in a 'branch', 'branches' sharing the same 'trunk' are grouped as linked list; Similarly, 'objects' belonging to the same 'branch' are also linked as list.

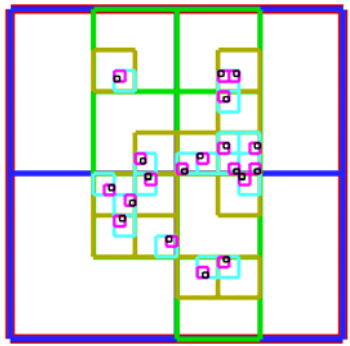 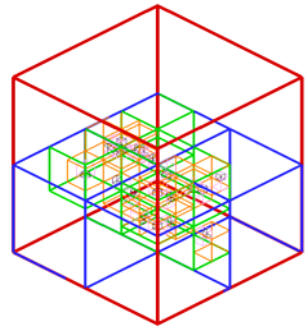

**Figure 1 Management region of SHT of two-dimensional discrete points**

**Figure 2 Management region of SHT of three-dimensional discrete points**

In practical applications the number of spatial objects may be variable. Therefore, the SHT is constructed dynamically. For the establishment of a 'tree' from an object, the typical procedure consists of two steps: (i) Get known the minimum resolution, i.e. the smallest edge length of cubes, σ; (ii) Use the center of object as the geometry center of cube. Check whether or not the cube can contain the object. If yes, the cube is a proper 'branch', then put the object into this branch; if not, the cubic length will be continued to double until the object can be contained, then create a 'branch' with the object placed in. Up to now, we have just established a 'tree' with only one 'branch' which contains one 'object'. Figure 3 shows the construction process of original tree, where a triangular object in two-dimensional space is used as an example.

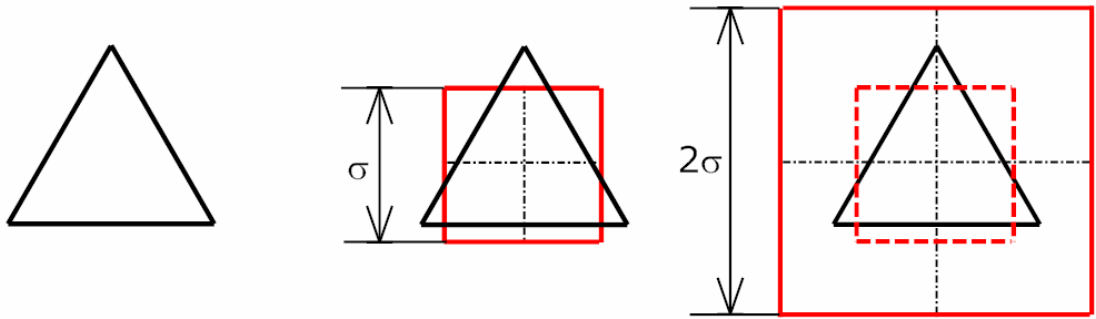

**Figure 3 Schematic for the construction of original tree**

For convenience, we use 'A' to represent already existed objects (possibly more than 1) on the tree. The algorithm for adding a new object B to the tree is as follows: (i) Establishing a new root. Check whether or not the object B can be contained by the old root. If not, then create a new root: Calculate the quadrant where the center of object B locates. Set the vertex of the old root which locates in this quadrant as the center of new root. In this way, the new root is the trunk of old one. Then, the old root becomes a sub-branch of the new one. Continue this process until the new roots can contain the object B. (ii) Placement of object B. Start from the new root. Compute the quadrant where the center of object B locates, with respect to current branch. Check whether or not the sub-branch of this quadrant contains object B. If not, the object B is placed in current branch; If yes, take the sub-branch as current branch. If the sub-branch does not exist, create it. Continue this process until a sub-branch is found or created which can contain object B. Figure 4 is schematic for the adding of a triangle to an existing tree.

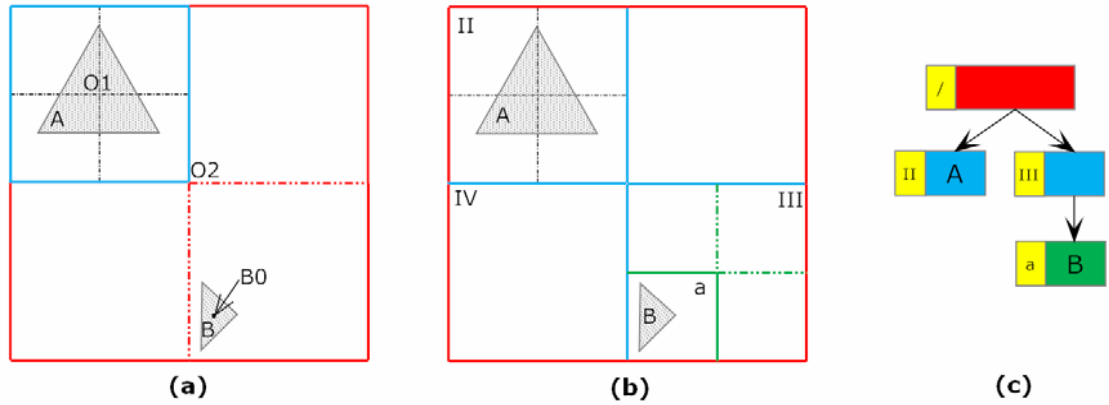

**Figure 4 Schematic for adding objects to tree. (a) generation of new root, (b) placing an object, (c) SHT corresponding to (b)**

The algorithm for removing an object from the tree is as follows: According to the link, pick up the branch containing the object. Remove the object from object-list corresponding to this branch. When a branch no longer contains objects and sub-branches, remove it. Enter its trunk, continue this process until all the useless branches are eliminated.

In dynamical algorithm of the SHT, except for adding a sub-branch or trunk of a branch, other operations have nothing to do with space dimension. The computer memory required by the SHT is approximately equal to $kN \ln N$, where $N$ is number of the objects. It is independent of space dimension. When the spatial dimension is higher or spatial objects are scatteredly distributed, the SHT can save a large quantity of memory compared with the background grid method. In addition, as SHT is dynamically constructed. The size of the system can dynamically increase or decrease with the addition or deletion of objects. This is a second obvious advantage over the traditional background grid method.

## 2  Fast searching algorithms based on SHT

When construct spatial geometry or determine neighboring relationship between objects, we need a fast search of objects satisfying certain conditions. The computational complexity of ergodic search is N. It is not practical to when dealing with a huge number of objects. For such cases, we need to develop fast searching algorithms.

By using the SHT we propose a fast searcher with computational complexity $\log N$. The basic idea is as below: We do not search directly objects, but check branches. Skip those branches without objects under consideration. Thus, searching is limited to a substantially small range. Depending on requirements of applications, we present two fast searching algorithms: conditional searching and minimum searching. The conditional searching is to search for objects meeting certain conditions. For example, to find out objects in a given area. Minimum searching is to search for an object whose function value is minimum. For example, to find a nearest object to a fixed point.

2.1 Conditional Search

The idea of conditional search is thus: First check whether or not a branch contains objects meeting some condition. If not, skip the branch. For example, to search for objects in a circle, one needs to assess whether or not the region of branch intersects with the circle. In this way, the searching is limited to the overlapped region of the branch and the circle.

Conditional searching is implemented using the stack structure. The steps are as follows: (i) Pushes the root into a stack A; (2) Pick out a branch b from the stack A; Check whether or not the objects in b satisfies the given conditions; Pick out the required objects; (3) Check each sub-branch of b; Push the branches satisfying the conditions into stack A; (4) Repeat (2)-(3) until the stack is empty.

Figure 5 shows the given circle and spatial division for managing planar triangles. Figure 6 is schematic for the SHT corresponding to figure 5, where '/' stands for the root. The process to find out objects in the given circle, showed in Fig. 7.

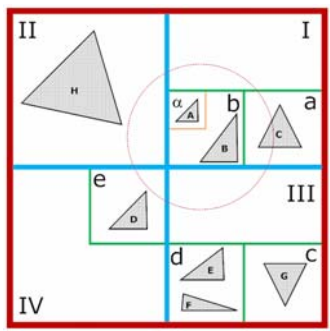
Figure 5 Distribution of planar triangles and the corresponding spatial division

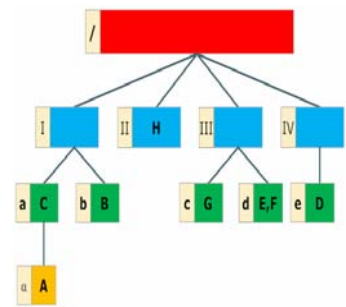
Figure 6 SHT corresponding to Figure 5

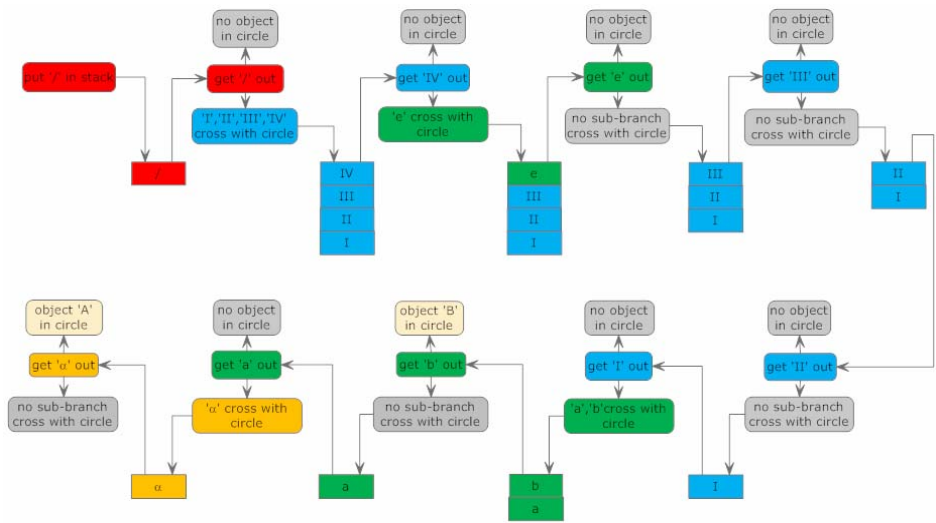

Figure 7 Flowchart for the fast search of objects in a circular area

The conditional searching is implemented by providing *conditional function* and *identification function*. The conditional function presents conditions which the objects should satisfy. The identification function is used to assess whether or not the region of a branch contain satisfying objects. It is clear that the validity of the algorithm is assured by the identification function. The more accurate the identification is, the fewer branches need to be searched. If identification status is always true, this searching algorithm goes back to an ergodic browser.

2.2 Minimum Search

For convenience of description, we define a few concepts. (i) *Range of a branch*: It means the range of the given function for objects in this branch. (ii) *B-R-branch*: It is a the new branch data structure composed of the branch itself and range of this branch. (iii) *Candidate B-R-branch*: It is the B-R-branch may be checked in the following procedure. It may contain objects whose functional values are minimum. In the minimum searching procedure, we must keep enough candidate B-R-branches. Some of them may be added or removed dynamically according to the need. In order to accelerate the searching speed, the candidate B-R-branches should be linked as a list. According to above definition, each B-R-branch has a range. So, each B-R-branch has a lower limit of its range. The B-R-branches in the list are arranged in such a way that their lower limits increase subsequently. Obviously, the B-R-branch with the smallest lower limit is placed on the head of the list. In addition, candidate objects and candidate values should be used to store the current objects with minimum values and the values themselves.

The idea of minimum searching is thus: By comparing the ranges of different branches, some branches can be excluded from the searching. The minimum searching algorithm is as follows: (i) The root and its range are merged as a B-R-branch; Add the B-R-branch to a candidate list named L; The candidate value V is set as positive infinity; The candidate object is set as null. (ii) Pick out a B-R-branch, for example, B, from candidate list L; Check the values of its objects. If the value of an object O is smaller than V, then, replace V with this value; at the mean time, set object O as candidate object. Remove the B-R-branches whose lower limit values are greater than V from the list L. (iii) Construct a B-R-branch Z for each sub-branch of B. If the minimum value of Z is larger than V, cancel Z; If the maximum value of Z is smaller than the minimum value of B-R-branch C in L, then all the B-R-branches behind C are remove from L; If the minimum value of Z is larger than the maximum value of a B-R-branch in L, cancel Z; Otherwise, insert Z into list L according to its lower limit. (4) Repeat steps (2) and (3) until the candidate list L is empty. The final candidate object is the required one.

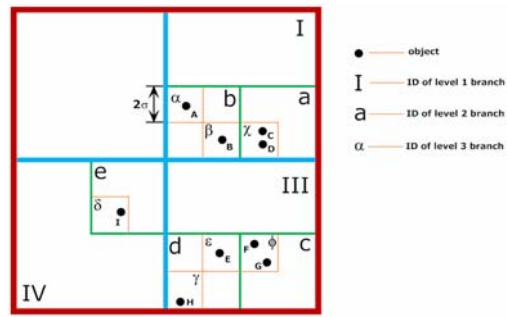

Figure 8 Planar point distribution and corresponding spatial division

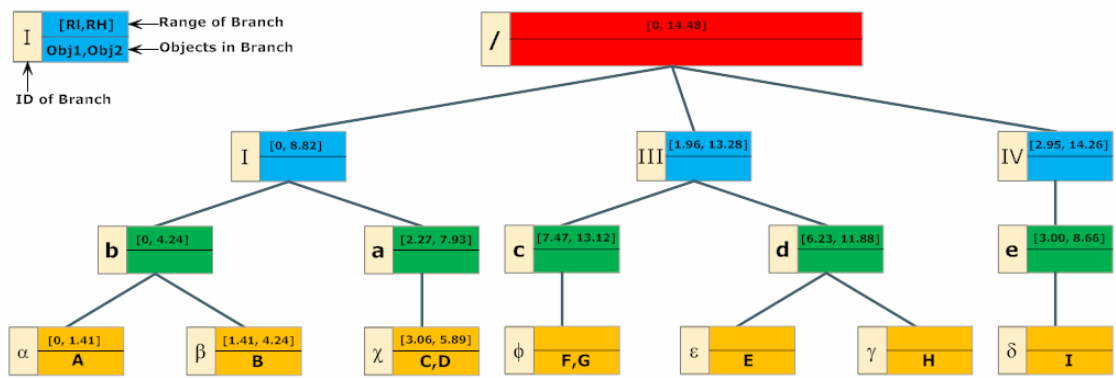

Figure 9 SHT corresponding to Figure 8

As an example for the applications of the proposed minimum search algorithm, we consider a case to find the nearest point to a fixed one from points in a plane. Figure 8 shows the distribution of planar points and spatial division. Figure 9 is the corresponding SHT. Suppose point A in Fig.8 is the given fixed point. To seek for the nearest point to it, the range of the branch is calculated by sphere evaluation method. Figure 10 shows the flow-chat.

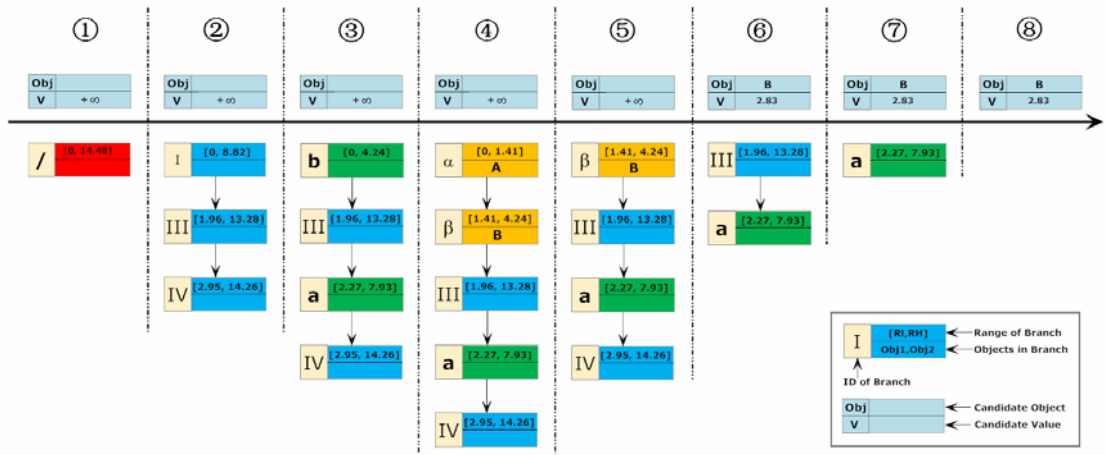

Figure 10 Schematic for the fast search of the nearest point to a given fixed one.

We can perform various minimum searches by providing different *value-finding function* and *range-evaluation function*. The value-finding function compute the value of object. The range-evaluation function assess the range of branch. The efficiency of minimum searching algorithm depends on the range-valuation function. The smaller the range given by the range-evaluation function, the faster the searching procedure. The worst range-evaluation function gives a range from $-\infty$ to $+\infty$. In such a case, the searching algorithm goes back to the ergodic browser. In the case with a large quantity of objects, one should use a good range-evaluation function to reduce the number of objects to be searched. But a good range-evaluation generally needs a large quantity of computations, which decreases also the global efficiency. We should find a balance between the two sides. Since the computation for sphere regions is more efficient than for cube ones, in complex minimum searching algorithms, circumspheres of cube are extensively used to evaluate the range of a branch.

## 3 Constructing Delaunay tetrahedron and triangle

Before constructing clusters, we divide the space using the given discrete points. We construct spatial geometrical structures by connecting given discrete points according to the Delaunay division approach. There are lots of algorithms to the construction of Delaunay tetrahedron in three-dimensional space or Delaunay triangle in two-dimensional space. The complexities of most algorithms are concentrated on the searching procedures. Here, we propose an algorithm based on the SHT. The algorithm is simple and intuitive. It is convenient to be extended to higher-dimensional space.

Main idea of the algorithm is as follows: when a new point is added to the formed Delaunay division structure, we should adjust the subdivision near the new point to meet the condition for Delaunay division. According to the definition of Delaunay division, if the new point is outside the circumsphere of a Delaunay simplex, this adding does not affect the Delaunay simplex. On the contrary, if it is inside a Delaunay simplex, it does affect the simplex. The simplex needs to be re-divided. Specifically, all the simplexes affected by the added new point are selected to form a complex. Each face of complex and the new point form a new simplex.

In three-dimensional space, the algorithm for construction of Delaunay tetrahedron from given discrete points is as follows: (i) Generate a sufficiently large tetrahedron to contain all discrete points; Record the center and radius of its circumsphere; Form a 'extended-tetrahedron' of the circumsphere; Add this extended-tetrahedron to SHT T (extended-tetrahedron SHT); (2) Pick out a discrete point P; Search in T for the extended-tetrahedra whose circumspheres contain P; Remove these extended-tetrahedra from T; Put the removed tetrahedra together to form a set named Q; (3) Add every surface of each tetrahedron in Q to SHT S which is a tree for the external triangular interfaces. Removes the surfaces that appears twice, because they are interfaces; Triangles in S constitute the external interface of Q ; (4) Pick out each face of S, together with point P, to construct new tetrahedron; Record the center and radius; Add the newly formed extended-tetrahedra to T; (5) Repeats steps (2) to (4) until all points are used out. The set of tetrahedra in T is just the required Delaunay division structure.

The algorithm includes two searches: one is for the circumsphere that contains a given point,

the other is for the external interface of Q. They both belong to the conditional search. When reduce to two-dimensional case, the algorithm keeps the same. We need only replace the tetrahedron with triangle and replace the triangular face with line. Figure 11 shows the procedure of adjusting space division due to the adding of a two-dimensional discrete point. Figure 12 shows the Delaunay triangle division of 20000 randomly distributed two-dimensional points. Figure 13 shows the Delaunay tetrahedron division of 20000 randomly distributed points in a three-dimensional sphere.

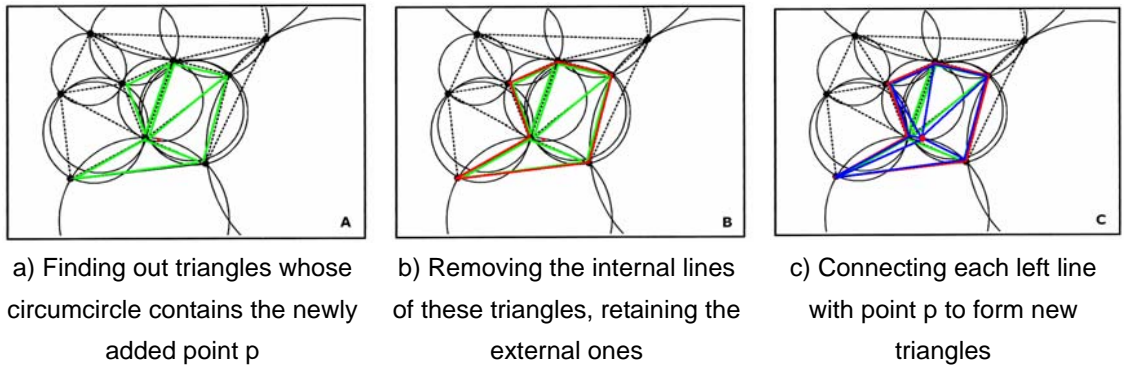

| a) Finding out triangles whose circumcircle contains the newly added point p | b) Removing the internal lines of these triangles, retaining the external ones | c) Connecting each left line with point p to form new triangles |

**Figure 11 Three steps to add a new point to a two-dimensional Delaunay division**

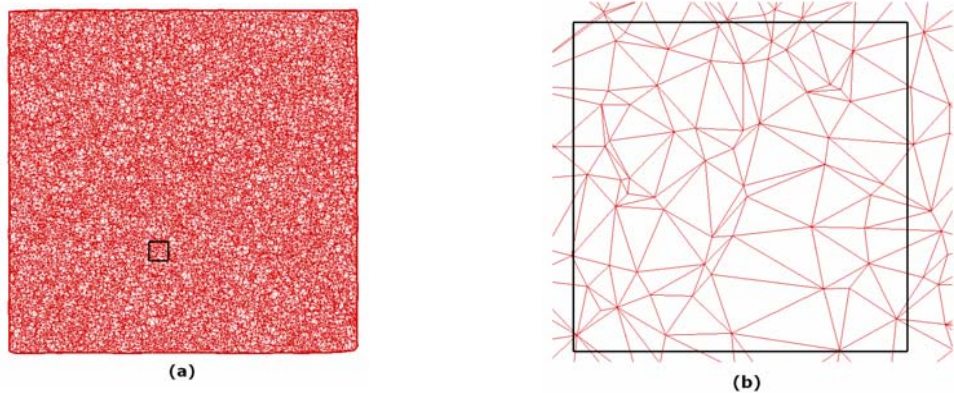

**Figure 12 Delaunay division constructed from randomly distributed discrete points in two-dimensional square area $[0,4]\times[0,4]$. Fig. (b) is the enlarged picture of the portion in the small black rectangle in Fig. (a).**

The algorithm can be easily extended to n-dimensional space. We need only replace the tetrahedron with n-simplex and replace the triangle with (n-1)-simplex. The circumsphere of n-simplex constructed from n+1 points, $\{\mathbf{r}_1, \mathbf{r}_2, \cdots, \mathbf{r}_{n+1}\}$, in n-dimensional space is used in the algorithm. The formula to calculate the center of circumsphere of n-simplex is $c_i = A_i / B$, where

$$A_i = \begin{vmatrix} 1 & -2\mathbf{r}_1 & \mathbf{r}_1^2 \\ 1 & -2\mathbf{r}_2 & \mathbf{r}_2^2 \\ \cdots & \cdots & \cdots \\ 1 & -2\mathbf{r}_{n+1} & \mathbf{r}_{n+1}^2 \\ 0 & \mathbf{e}_i & 0 \end{vmatrix}, \quad B = \begin{vmatrix} 1 & -2\mathbf{r}_1 & \mathbf{r}_1^2 \\ 1 & -2\mathbf{r}_2 & \mathbf{r}_2^2 \\ \cdots & \cdots & \cdots \\ 1 & -2\mathbf{r}_{n+1} & \mathbf{r}_{n+1}^2 \\ 0 & \mathbf{0} & 1 \end{vmatrix},$$

$\mathbf{e}_i$ is the unit vector of the i-th derection.  The radius of the circumsphere is $\sqrt{(\mathbf{c}-\mathbf{r}_1)^2}$.

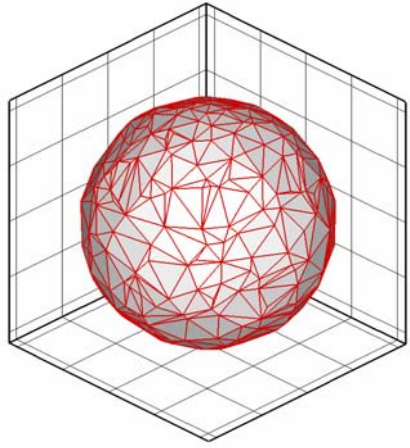

**Figure 13 Delaunay division constructed from randomly distributed**

**discrete points in three-dimensional spherical region**

## 4  Cluster construction and analysis method

For the discrete points in space, there is no strict cluster structure. If the discrete points are considered as objects, such as molecular ball, lattice or grid, then, the objects can be connected to form clusters. The average size of these assumed objects is the revolution of clusters to be constructed with discrete points. The constructing of clusters is very simple. A cluster is formed by connecting all points whose distance in between is less than the revolution length.

After the construction of Delaunay division for given discrete points, remove the lines whose lengths are greater than the revolution length. The remained spatial structure may have various dimensions. According to connectivity, the structures that are not connected to each other can be decomposed into different clusters. Each cluster may have also structures with various dimensions. For example, structure consisted of two triangles with a common side, or structure formed by a triangle and a tetrahedron, etc. In physical problems, the structures with high-dimensional measure play a major role on the system. Generally, we need analyze only clusters with the maximum dimensions.

    The cluster construction algorithm consists of three parts. Preparation part: (i) Construct Delaunay tetrahedra from given discrete points. The corresponding SHT is notated as t. (ii) Remove the tetrahedrons whose length is greater than given resolution from t. Single cluster construction part: (iii) Remove tetrahedron T from t if such a T still exists. Create a new cluster named C. Initialize the body tree C->t and face tree C->s as null. Add T to the body tree C->t. Add each triangle faces to a triangle tree named i. (iv) Pick out a triangle face S from i. Search tetrahedron Y containing face S from t. If found, add Y to tree C->t and add all faces of Y to tree i. Two faces with opposite directions will annihilate if they meet with each other during the adding procedure. If not found, add S to the tree C->s. (v) Repeat step (iv) until the tree i becomes null. Construct all the clusters: (vi) Add constructed cluster C to a tree for clusters named c. Repeat the process of constructing single cluster, add the new cluster to c, until t becomes null. The algorithm for adding a face S to the tree i is as follows. Search and check if a face with the opposite direction of S exists in the tree i. If exists, remove it from the tree i. If not exist, add S to the tree i. Up to now, all the constructed clusters are put to the tree for clusters c. For each cluster C in the tree c, all tetrahedron elements are placed on the body tree C->t, all the surface triangles are placed on the tree for faces C->s. Figures 14 and 15 shows respectively the clusters constructed with random points in two-dimensional and three-dimensional space.

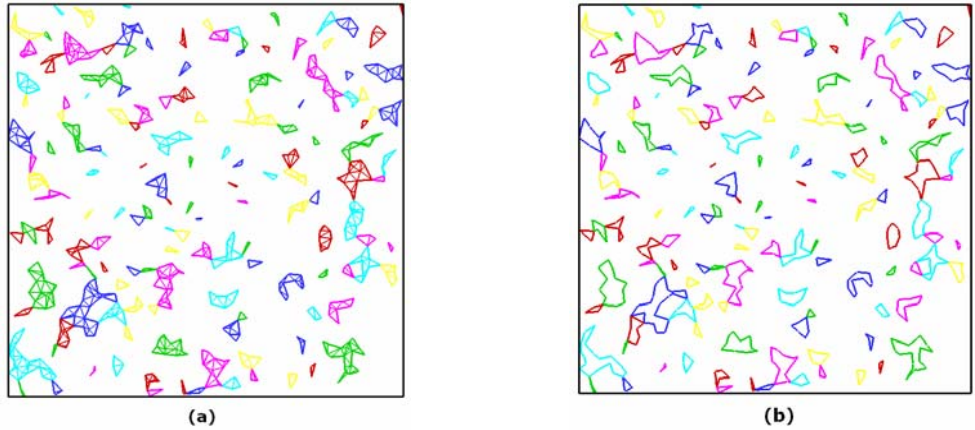

**Figure 14 Cluster structure formed from 1000 random discrete points in two-dimensional square area $[0,1]\times[0,1]$. Fig. (a) is for a cluster. Fig. (b) is for the corresponding cluster boundary.**

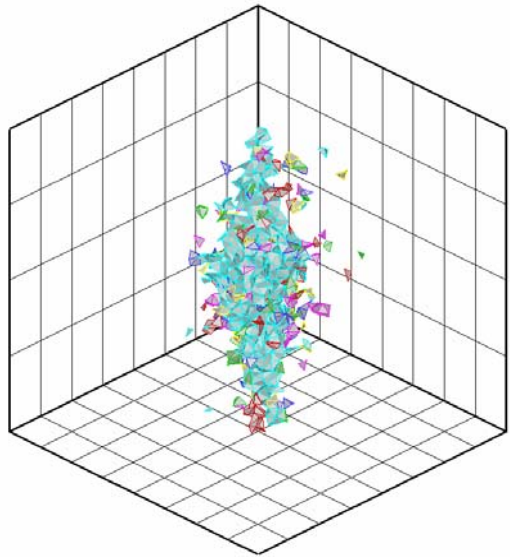

**Figure 15 Cluster structure formed form 5000 three-dimensional random discrete points**

The algorithm is also applicable to n-dimensional discrete points. We need only replace the tetrahedron with n-simplex and replace the triangle surface with (n-1)-simplex. For space with a dimension higher than three, the number of neighboring points and the connectivity, as well as the number of n-simplex, grow rapidly with the dimension. So, the required memory increases quickly. Delaunay division can be constructed partition by partition. The main skill in this algorithm is that the space is partitioned according to the main branches of SHT, points in each partition are added sequentially. After the completion of adding all points in a partition, we need delete the n-simplex satisfying two conditions: (i) its external circumsphere is in the completed partition, (ii) at least one side is longer than the given resolution.

**6 Conclusions and discussions**

We propose a new method for managing objects and fast searching in arbitrary dimensional space.  Based on this, algorithms for the constructing Delaunay simplex and spatial clusters are presented. The applications to two- and three-dimensional discrete points validate the method and show obvious advantages.

The proposed SHT can be easily used to manage and search objects with various locations, sizes or even shapes. This management method can be widely used in many fields. As an example, in the finite-element method, due to the complexity in calculating relative positions of elements and in the searching for them, only the cases with simple shapes and single-sizes are extensively studied. With the proposed SHT, it is easy to search adjacent relationship between objects with various sizes and shapes. Therefore, the SHT can substantially simplify simulations with the movements of complex objects.

Compared with previous methods, the proposed SHT and the two fast-searching algorithms based on it are established on a more abstract framework. It has potential extensibility to various fields. Experimental data can also be treated with under the same SHT after parameterization of physical quantities. Then, fast searching can be realized according to a similar algorithm. In addition, according to the needs, an object can be simultaneously placed in several SHTs. This is equivalent to set indexes of several properties. Therefore, fast searching for various properties can be easily implemented.

```
Refences:
```

1    Ament P. Effects of ionization gradients on Inertial-Confinement-Fusion capsule hydrodynamics stability. Phys Rev Lett, 2008, 101: 115004-1—4.
2    de Vries1 P C, Hua M D, McDonald D C, et al. Scaling of rotation and momentum confinement in JET plasmas, Nucl Fusion 2008, 48: 065006-1—18.
3    Bowler B P, Waller W H, Megeath S T, et al. An infrared census of star formation in the horsehead nebula. The Astronomical Journal 2009, 137: 3685-3699
4    L. Hernquist. Hierarchical N-body methods. Comput. Phys. Commun. 1988,48: 107-115.
5    Makino J. Vectorization of a treecode. J Comput Phys, 1990, 87: 148-160 .
6    Hidaka Y, Choi E. M., Mastovsky I., et al. Observation of large arrays of plasma filaments in air breakdown by 1.5-MW 110-GHz gyrotron pulses. Phys. Rev. Lett. 2008,100:035003

7   Nogaret T, Rodney D, Fivel M, et al. Clear band formation simulated by dislocation dynamics: Role of helical turns and pile-ups. J Nucl Mater, 2008, 380: 22-29.

8   Kotsiantis S B and Pintelas P E. Recent advances in clustering:a brief survey. WSEAS Trans. Inform. Sci. Appl. 2004, 1:73-81.

9   Fan Y J, Iyigun C, Chaovalitwongse W A. Recent advances in mathematical programming for classification and cluster analysis. CRM Proceedings and Lecture Notes, 2008, 45: 67-93.

10  Chazelle B, Devillers O, Hurtado F, et al. Splitting a Delaunay triangulation in linear time. Algorithmica 2002, 34: 39-46.

11  Clarkson K L, Varadarajan K. Improved approximation algorithms for geometric set cover. Discrete & Computational Geometry, 2007, 37(1): 43-58

12  Black P E(ed.), Entry for data structure in Dictionary of Algorithms and Data Structures. U.S. National Institute of Standards and Technology. 15 December 2004

13  Knuth D E. The Art of Computer Programming. Vol. 3: Sorting and Searching, Addison Wesley Longman Publishing Co., Inc., Redwood City, CA, 1998.

14  Barnes J and Hut P. A hierarchical $O(N \ln N)$ force-calculation algorithm, Nature (London) , 1986, 324:   446-449.

15  Pfalzner S and Gibbon P. Many Body Tree Methods in Physics, Cambridge University Press, New York, 1996.

16  Nam B, Sussman A. A comparative study of spatial indexing techniques for multidimensional scientific dataset. Proceedings of the 16th International Conference on Scientific and Statistical Database Management (SSDBM'04), p.171, June 21-23, 2004.